# Exploring Urban Form through OpenStreetMap Data: A Visual Introduction

Geoff Boeing
Department of Urban Planning and Spatial Analysis
University of Southern California

Abstract

This chapter introduces OpenStreetMap—a crowd-sourced, worldwide mapping project and geospatial data repository—to illustrate its usefulness in quickly and easily analyzing and visualizing planning and design outcomes in the built environment. It demonstrates the OSMnx toolkit for automatically downloading, modeling, analyzing, and visualizing spatial big data from OpenStreetMap. We explore patterns and configurations in street networks and buildings around the world computationally through visualization methods—including figure-ground diagrams and polar histograms—that help compress urban complexity into comprehensible artifacts that reflect the human experience of the built environment. Ubiquitous urban data and computation can open up new urban form analyses from both quantitative and qualitative perspectives.

# Introduction

New technologies have recently changed how we can see, understand, and plan the urban form. Cognitive science and biometric tools such as eye-tracking technology help researchers study human psychological and physiological experiences in the built environment to advance evidence-based urban design (Sussman and Hollander, 2015; Sussman and Ward, 2016). New spatial information platforms allow urban scholars and practitioners to explore the patterns, textures, and connectivity of the urban fabric—while empowering public participation and engagement (Boeing, 2018; Evans-Cowley and Hollander, 2010). Meanwhile, the Smart Cities paradigm of urban governance seeks to monitor and control cities via ubiquitous sensing and automatically harvested data, both to track how humans move through space and to respond to their behavior (Albino et al., 2015; Batty et al., 2012).

Yet too often, urban technologies are used merely for top-down monitoring, optimization, and control. Rather than just considering urban livability as a naive optimization problem, city planners could instead use big data to enrich socio-political processes of community advocacy, consensus-forming, and public decision-making (Goodspeed, 2015). Or, planners could use new data and tools to investigate sampling biases in user-generated content, adjust for over-representation of certain groups, and foreground the social experiences of marginalized populations (Boeing et al., 2020; Kang et al., 2019). User-generated spatial data can even help us introspectively unpack planning and design histories and the spatial logics they manifest, which in turn shape human behavior and experience.

This chapter explores a growing source of big data on spatial infrastructure, to reflect on how the built environment constrains and shapes the human experience in urban space. Street networks are perhaps the paradigmatic example of such infrastructure. Urban planners model these networks to investigate trips and traffic, explore urban planning and design histories, and better understand the psychology of human navigation and wayfinding in the built environment. This chapter introduces OpenStreetMap (OSM)—a worldwide mapping community and online geospatial data repository—and tools to work with and visualize its data. OSM provides a freely available, high quality source of data on street networks and other infrastructure worldwide (Jokar Arsanjani et al., 2015). We will discuss what it is, how to use it, and how it can help us critique and improve the urban experience.

# Background and Methods

### What Is OpenStreetMap?

First launched in 2004, OSM is a wiki-style, crowd-sourced, worldwide mapping project and geospatial data repository with good coverage and quality (Barron et al., 2014; Basiri et al., 2016; Corcoran et al., 2013; Girres and Touya, 2010; Haklay, 2010; Sehra et al., 2019;

Zielstra et al., 2013). Think of it as Wikipedia meets Google Maps. Volunteers provide some editorial oversight of edits, but anyone may edit the map using tools such as ESRI's ArcGIS Editor for OpenStreetMap. OSM contains data on streets and highways, transit systems, building footprints, parks and plazas, pedestrian and bicycle infrastructure, political boundaries, and more (though non-road coverage varies somewhat around the world).

To date, over 1 million different users have contributed content including 5.6 billion nodes (i.e., geospatial points), 600 million ways (i.e., geospatial lines), and related descriptive data. It is not a perfect data source: researchers estimate that >95% of OSM contributors are male, suggesting possible content creation biases (Schmidt & Klettner, 2013). Nevertheless, OSM is public, free, global, and open to anyone to contribute new data (such as streets, building footprints, or points of interest) if they notice something missing.

## How to Work with OpenStreetMap Data

OSM offers application programming interfaces (APIs) which allow anyone to write snippets of code to query its databases and return spatial data. But by itself, OSM (and its various APIs) can be challenging to work with, particularly for planning and design practitioners without strong technical backgrounds. Its raw data do not lend themselves automatically to urban form/network analysis and its custom query languages can be cumbersome for scripting.

Fortunately, there is an easier way. In this chapter we explore urban form and street networks using OSMnx, a simple software package for OSM spatial data collection, modeling, analysis, and visualization (Boeing, 2017). OSMnx is free, open-source, and fully documented. It allows researchers and practitioners to easily download street network, building, and amenity data for any study site in the world, then automatically construct them into street network models or spatial dataframes for built-in visualization and statistical analysis. OSMnx allows users to download spatial data from OSM for any study site boundary in the world. This unlocks new ways of engaging with this massive repository of global spatial information to see and understand the urban form, as we will demonstrate momentarily.

But first, a quick technical note for the sake of clear terminology: OSMnx's street network models are what are called "nonplanar directed graphs." The word *graph* is just a synonym in mathematics for *network*, and in our case refers to a model (i.e., a representation in your computer's memory) of a real-world network. Graphs are made up of nodes and edges. In the case of a city street network, graph *nodes* represent intersections and dead-ends, and graph *edges* represent the street segments linking them (Barthelemy, 2011; Boeing, 2017; Cardillo et al., 2006; Marshall et al., 2018). A *directed* graph means that all the connections in the network point in a single direction, for instance from node A to node B. This allows us to model one-way streets, while two-way streets just have links pointing in both directions. *Nonplanar* just means that our models can exist in three dimensions rather than only in a two-dimensional plane. This may seem obviously important, considering the prevalence of three-dimensional overpasses and underpasses in many cities, but you may be surprised to learn that many urban street network datasets and tools are *planar* instead—owing to the legacy of two-dimensional cartography and planar graph research in mathematics and physics. Given the realities of urban form, OSMnx uses nonplanar models.

With that quick survey of terminology in the bag, let us consider some practical uses of OSM and its massive data repository.

**What Is OpenStreetMap Useful For?**

OSM's data can tell us about urban form and spatial patterns. Modeling an urban street network, we can simulate trips to explore commuting patterns and travel demand. We can analyze the structure of the network to understand resilience and vulnerability—that is, where is the infrastructure more brittle and prone to fail during an extreme event like flooding or earthquakes? OSM data let us look across cities—even countries—to explore urban patterns and configurations from different decades or cultures, and think about what different design paradigms mean for urban living.

Many urban researchers, planners, and designers have used OSM data and OSMnx accordingly to study and improve the built environment. For instance, Hernández-Hernández et al. (2019) use them to explore commuter routes in a study of motorists' emotions and expressions of anger while driving in Mexico City. Natera Orozco et al. (2020) calculate quality-of-life indicators in Budapest, using OSMnx to model its walkable network and local amenities. Liu et al. (2020) model Beijing's walkable street network with OSMnx to explore spatial patterns of residents' daily leisure activities. Natera Orozco et al. (2019) model "as-is" bicycle networks to demonstrate how cities can design small, targeted infrastructure changes to significantly increase connectivity and directness. The mobile crowd-sensing platform CrowdSenSim uses OSMnx to simulate urban environments (Montori et al., 2019; Tomasoni et al., 2018) and the transportation planning company Remix developed its street design platform—now deployed in hundreds of cities worldwide—using OSMnx to model city streets. Padgham et al. (2019) use OSM data to demonstrate siting hospitals for faster time-sensitive access, such as for stroke treatment. OSM's street and urban form data were widely used in the humanitarian response to 2010's catastrophic Haitian earthquake (Zook et al., 2010).

Here we explore urban form visually using OSM data, to illustrate the context and history of urban planning and design in different places. In particular, we reflect on how the urban form shapes urban living and the urban experience. Cities have changed drastically over the past century. Planners reorganized urban space around the logic of the automobile, gutting central cities in favor of sprawling mobility and settlement patterns. This has had significant ramifications for pedestrian access, safety, health, comfort, culture, and wayfinding. We explore these urban forms and histories using two primary visualization methods with OSM data.

The first method uses OSMnx to produce figure-ground diagrams of street networks and building footprints, for the illustration of urban design and planning decisions. Shanken (2018) demonstrates how urban planners throughout history have employed a constellation of visual methods to analyze spatial information and represent the city. This representational visual culture was exemplified by Giambattista Nolli's 18th century ichnographic study of Rome, producing the famous figure-ground Nolli Maps of the urban fabric (Hwang and Koile, 2005; Verstegen and Ceen, 2013). Two centuries later, this visual methodology was operationalized by Allan Jacobs' (1995) comparative study of dozens of urban street networks around the world and their impacts on human navigation, cognition, and experience. The heart of Allan Jacobs's (1995) classic book on street-level urban form and design, *Great Streets*, features dozens of hand-drawn figure-ground diagrams in the general style of Nolli maps. We adapt this visualization methodology to a computational, big data workflow to similarly depict one square mile of multiple cities' street networks, to compare their street networks and urban forms.

The second method uses polar histograms of street orientations to uncover planners' spatial ordering of the built environment (Boeing, 2019a; Gudmundsson and Mohajeri, 2013; Mohajeri et al., 2013a, 2013b; Mohajeri and Gudmundsson, 2014, 2012). To generate these visualizations, we calculate the compass bearings of all the street segments in 25 world cities, then visualize them with a polar histogram in which the bars' directions represent 10° bins around the compass, and the bars' lengths represent the relative frequency of street segments that fall in each bin (see Boeing, 2019a for full details). This produces a visual representation of the extent to which a street network follows the spatial ordering logic of a single grid versus having streets oriented more evenly in all compass directions.

## Results and Discussion

### Urban Circulation Systems and Spatial Logics

Figure 1 shows one-square-mile figure-ground diagrams from 12 cities around the world. At the top-left, Portland, Oregon and San Francisco, California typify the late nineteenth century orthogonal grid (Cole, 2014; Marshall et al., 2015; Southworth and Ben-Joseph, 1997, 1995). Portland's famously compact, walkable, 200-foot × 200-foot blocks are clearly visible but its grid is interrupted by the Interstate 405 freeway which tore through the central city in the 1960s (Mesh, 2014; Speck, 2012). In the middle-left, the business park in suburban Irvine, California demonstrates the coarse-grained, modernist, auto-centric form that characterized American urbanization in the latter half of the twentieth century (Hayden, 2004; Jackson, 1985; Jacobs, 1995).

In stark contrast, Rome has a more fine-grained, complex, organic form evolved over millennia of self-organization and urban planning (Taylor et al., 2016). Because we represent all of these street networks here at the same scale—one square mile—it is easy to compare the qualitative urban patterns in these different cities to one another. Contrast the order of the orthogonal grid in San Francisco and the functionalist simplifications of Irvine to the messy, complex mesh of pedestrian paths, passageways, and alleys constituting the circulation network in central Rome. Imagine the pedestrian experience walking through Rome versus along the periphery of Irvine's superblocks. Consider the design principles manifested in the urban fabric—and their implications for human cognition and comfort from the perspectives of biophilia, thigmotaxis, and prospect-refuge theory (Sussman and Hollander, 2015).

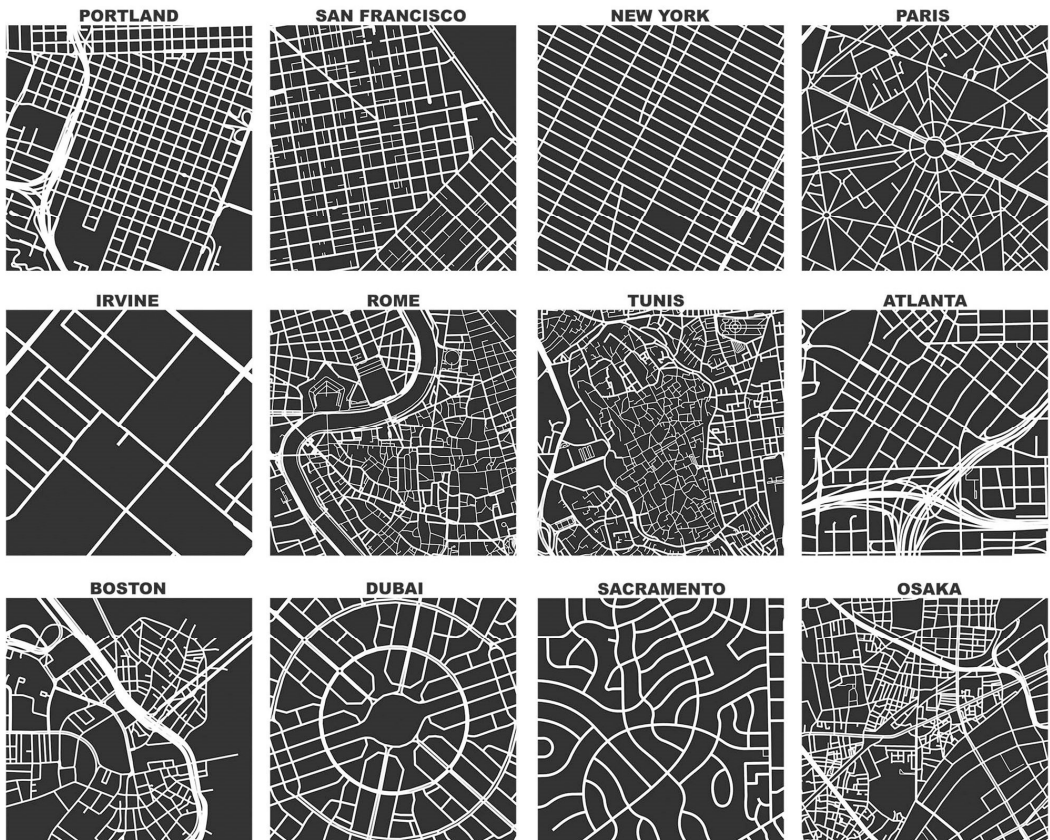

**Figure 1**. One square mile of each city's street network. The consistent spatial scale allows us to easily compare different kinds of circulation networks and urban forms in different kinds of places.

At the top- and middle-right, we see New York, Paris, Tunis, and Atlanta. Midtown Manhattan's rectangular grid originates from the New York Commissioners' Plan of 1811, which laid out its iconic 800-foot × 200-foot blocks approximately 29 degrees off true North (Ballon, 2012; Koeppel, 2015; Marcuse, 1987). Broadway weaves diagonally across it, revealing the path dependence of the old Wickquasgeck Trail's vestiges, which Native American residents used to traverse the length of the island long before the first Dutch settlers arrived (Holloway, 2013; Shorto, 2004). This historical route still shapes human wayfinding in the city today.

At the center of the Paris square mile lies the Arc de Triomphe, from which Baron Haussmann's streets radiate outward as remnants of his massive demolition and renovation of nineteenth century Paris (Hall, 1996). The spatial signatures of Haussmann's design project can clearly be seen via network analysis through the redistribution of network centralities and block sizes (Barthelemy et al., 2013), essentially restructuring urban centrality and travel behavior while drastically reshaping the human experience of moving through Paris. At the center of the Tunis square mile lies its Medina, with a complex urban fabric that evolved over the middle ages (Kostof, 1991; Micaud, 1978). Finally, Atlanta is typical of many American downtowns: coarse-grained, disconnected, and surrounded by freeways (Allen, 1996; Fishman, 2011; Grable, 1979; Jackson, 1985; Kruse, 2007; Rose, 2001).

The bottom row of Figure 1 shows square miles of Boston, Dubai, Sacramento, and Osaka. The central Boston square mile includes the city's old North End—beloved by Jane Jacobs (1961) for its lively streets and engaging human-scaled complexity, but previously cut-off from the rest of the city by the Interstate 93 freeway. This freeway has since been undergrounded as part of the "Big Dig" megaproject to alleviate traffic and re-knit the urban fabric (Flyvbjerg, 2007; Robinson, 2008). The Dubai square mile shows Jumeirah Village Circle, a master-planned residential suburb designed in the late 2000s by the Nakheel corporation, a major Dubai real estate developer (Boleat, 2005; Haine, 2013; Kubat et al., 2009). Its street network demonstrates a hybrid of the whimsical curvilinearity of the Garden Cities movement and the ordered geometry of modernism.

The Sacramento square mile depicts its northeastern residential suburb of Arden-Arcade and demonstrates Southworth and Ben-Joseph's "warped parallel" and "loops and lollipops" design patterns of late twentieth century American urban form that reordered cities around the logic of the automobile (Southworth and Ben-Joseph, 1997). Finally, the Osaka square mile portrays Fukushima-ku, a mixed-use but primarily residential neighborhood first urbanized during the late nineteenth century. Today, the freeway we see in the upper-right of this square mile infamously passes through the center of the high-rise Gate Tower Building's fifth through seventh floors (Yakunicheva, 2014).

To compare urban patterns in different kinds of places, these visualizations depict modern central business districts, ancient historic quarters, twentieth century business parks, and suburban residential neighborhoods. The cities they represent are drawn from across the United States, Europe, North Africa, the Arabian Peninsula, and East Asia. Yet street network patterns also vary greatly *within* cities: Portland's suburban east and west sides look different than its downtown, and Sacramento's compact, grid-like downtown looks different than its residential suburbs—a finding true of many American cities (Boeing, 2020). A single square-mile diagram thus cannot be taken to be representative of broader scales or other locations within the municipality. These visualizations, rather, show us how different urbanization patterns and paradigms compare at the same scale, using automatically harvested user-generated data. This can serve both as a practitioner's tool for investigating the physical outcomes of planning and urbanization, as well as a tool for communicating design in a clear and immediate manner to laymen—leveraging spatial data to improve collaboration and co-governance.

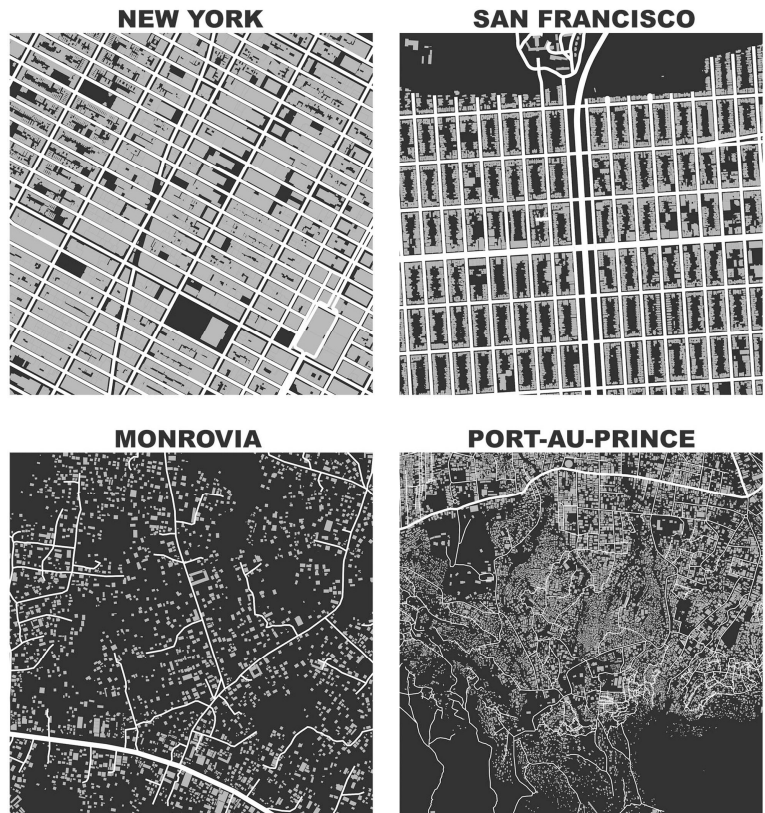

**Figure 2**. One square mile of each city's street network and building footprints. The consistent spatial scale allows us to easily compare the urban form in different kinds of places.

## Buildings and the Urban Fabric

These use cases can be seen even more clearly when we use OSMnx to visualize street networks along with building footprints, as shown in Figure 2. At the top-left, we see the densely-built form of midtown Manhattan, with large buildings filling most of the available space between streets and providing a strong sense of enclosure in the public rights-of-way. Within this square mile, there are 2,237 building footprints with a median area of 241 square meters. At the top-right, we see the medium-density perimeter blocks of San Francisco's Richmond district, just south of the Presidio. Here the building footprints line the streets while leaving the centers of each block as open space for residents—the enclosure is privatized. Within this square mile, there are 5,054 building footprints with a median area of 142 square meters.

The bottom two images in Figure 2 reveal an entirely different mode of urbanization by visualizing the slums of Monrovia, Liberia and Port-au-Prince, Haiti. These informal settlements are much finer-grained and are not structured according to the centralized geometric logic of the American street grids in the top row. Monrovia's square mile contains 2,543 building footprints with a median area of 127 square meters. Port-au-Prince's square mile contains 14,037 building footprints with a median area of just 34 square meters.

OSM data and OSMnx provide practitioners an easy-to-use tool to analyze and visualize the built environment for better planning and design. For instance, the data in Figure 2 could help designers and residents in Monrovia and Port-au-Prince collaboratively plan how to integrate new formal circulation networks in these informal settlements with minimal disruption to the existing urban fabric, homes, and livelihoods (Brelsford et al., 2019, 2018; Masucci et al., 2013).

## Modernist Inversion of Spatial Order

Visualizing this spatial information can reveal modernism's inversion of traditional urban spatial order and the state assertion of power over social life, culture, and human experience (Holston, 1989; Vale, 2008). In pre-industrial cities (as seen in Figure 3), the "figure" dominates the "ground" as the diagram displays scattered open space between buildings. This exemplifies a traditional sense of enclosure in human habitats that supports the experience of moving through urban space (Sussman and Hollander, 2015). But in modernist cities, the "ground" dominates the "figure" as only a few scattered buildings are positioned as sculptural elements across the landscape's void.

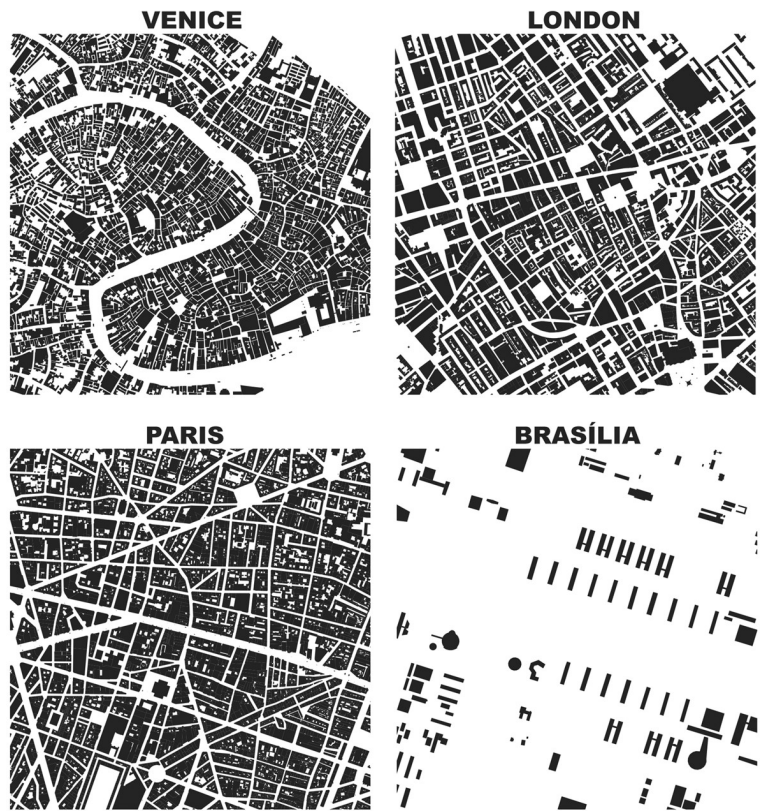

**Figure 3**. One square mile figure-ground diagrams of building footprints in the city centers of Venice, London, Paris, and Brasília illustrate the modernist inversion of traditional urban spatial order and the loss of sense of enclosure in modernist cities.

This modernist design paradigm sought to open up the dense and messy urban fabric with towers-in-the-park, spacing, and highways (Fishman, 2011; Jacobs and Appleyard, 1987)—drastically changing the cognitive, emotional, and psychological experience of the city. This phenomenon is clearly seen in Brasília, the modernist capital of Brazil, designed as a planned city in the 1950s by Lúcio Costa, Oscar Niemeyer, and Roberto Burle Marx (Figure 3). The structural order of the city suggests "an ordering of social relations and practices in the city" (Holston, 1989, p. 125). These figure-ground diagrams provide a spatial data-driven way to qualitatively study the urban form that shapes human travel behavior and social psychology.

## Street Network Orientation

The polar histograms in Figure 4 offer another perspective on this structural ordering of the city. For example, in Manhattan's polar histogram we can see the spatial order produced by its dominant orthogonal grid (cf. Figure 1) as its street bearings are primarily contained in four bins, offset from true north. Higher-entropy (i.e., more-disordered) orderings of street orientations can be seen in the other boroughs.

In Figure 5 we see polar histograms of 100 cities around the world. While some street networks in modern cities in the US, Canada, Australia, and China demonstrate similar low-entropy grids, more of these cities exhibit higher entropy. That is, their streets are oriented more evenly in all compass directions rather than following the spatial ordering logic of one or two consistent grids. The spatial signature of the street grid in these data rises to the forefront in Figure 5 in cities like Chicago and Toronto, while cities like Rome and São Paulo demonstrate more organic patterns with less unified, rigid, geometric planning. Consider an older American city like Boston (Figure 6). Although it features a grid in some neighborhoods like the Back Bay and South Boston, these grids tend to not be aligned with one another, resulting in an amalgam of competing orientations. Furthermore, these grids are not ubiquitous and Boston's other streets wind in various directions. In many parts of town, if you are traveling north and then take a right turn, you might know that you are immediately heading east, but it is difficult to know where you are eventually heading in the long run. This can make navigation difficult, as it does not adhere to a consistent, predictable spatial pattern—not because Boston apocryphally "paved over its cow paths" but because of its age, terrain, and annexation of various independent towns.

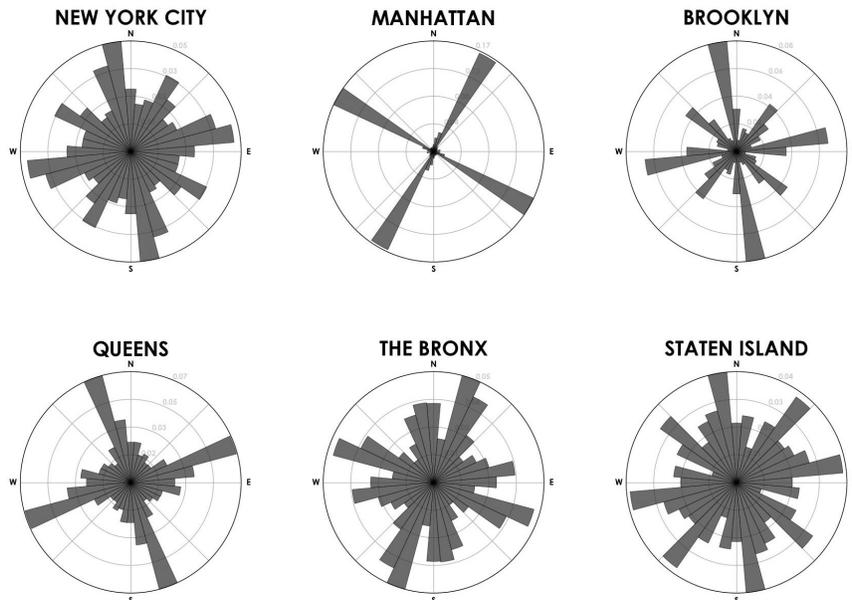

**Figure 4**. Polar histograms of the street orientations in New York City and its five constituent boroughs.

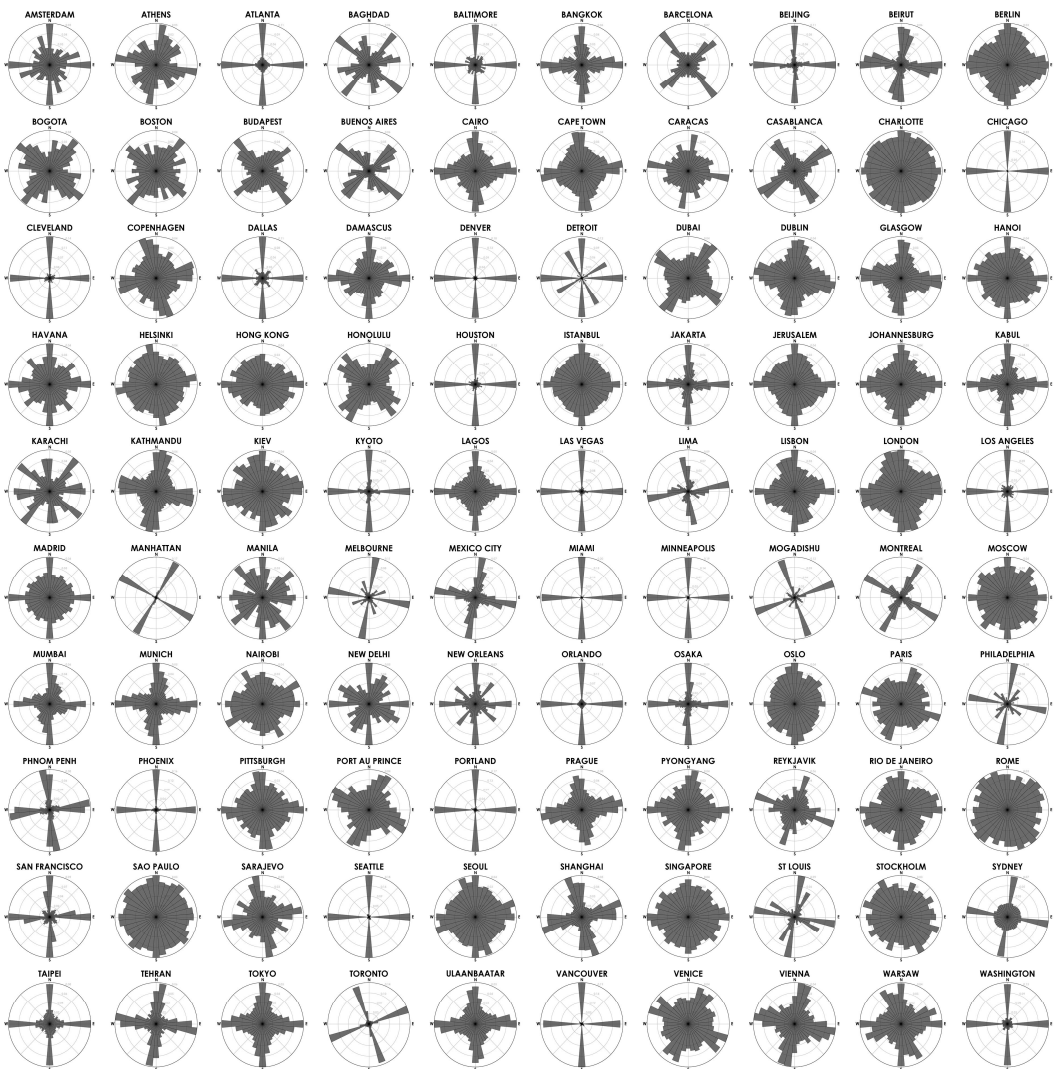

**Figure 5**. Polar histograms of the street orientations in 100 cities around the world.

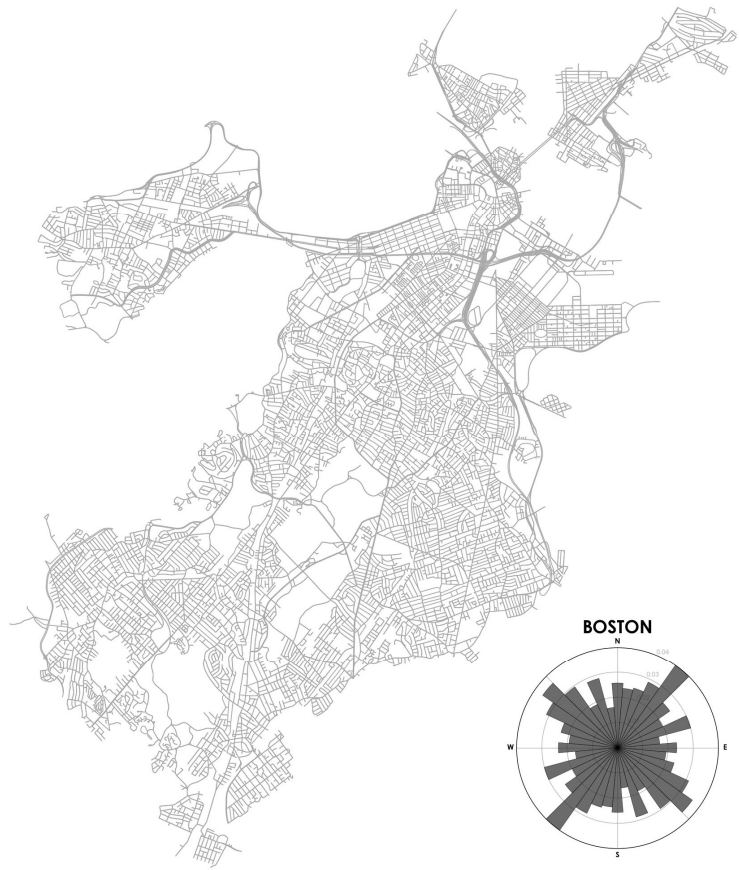

**Figure 6**. Boston's street network and corresponding polar histogram of street orientations.

Sixty years ago, Kevin Lynch (1960) defined *legible* cities as those whose patterns lend themselves to coherent, organized, recognizable, and comprehensible mental images. These characteristics help us organize city space into cognitive maps for wayfinding and a sense of place (Lynch, 1984; Lynch and Rodwin, 1958). But what Boston lacks in straightforward circulation patterns, it makes up for with other Lynchian elements (e.g., paths, edges, districts, nodes, landmarks) that help make it an *imageable* city for locals and visitors. From Lynch (1960, p. 10):

> "A highly imageable city in this peculiar sense would seem well formed, distinct, remarkable; it would invite the eye and the ear to greater attention and participation… The concept of imageability does not necessarily connote something fixed, limited, precise, unified, or regularly ordered, although it may sometimes have these qualities.

Nor does it mean apparent at a glance, obvious, patent, or plain. The total environment to be patterned is highly complex, while the obvious image is soon boring and can point to only a few features of the living world."

## Conclusions

Through the tools of computer science, data science, and visualization, spatial information allows us to see how urban evolution, planning, design, and millions of individual human decisions shape how cities organize and order space according to various spatial logics. This chapter introduced figure-ground diagrams and polar histograms as methods of hybrid quantitative-qualitative analysis of urban patterns and the human experiences they shape. These visualizations reveal the texture, grain, and spatial logic of different cities around the world. Compressing the dense complexity of information inherent to cities, they offer a streamlined view of the urban fabric and how the circulation system connects it. The figure-ground diagrams allow us to compare across places at the same scale to visualize similarities and differences. The polar histograms compress the complexity of street network orientation into simple plots that reveal the spatial order of the city's streets.

      These visualization tools and techniques can help planners convey comparative urban form to laypersons. They can destigmatize density and enclosure, and explain how connectivity and texture vary across cities. Finally, they simplify complicated urban planning and data science concepts to make them more approachable to engage members of the public. This can provide a comprehensive understanding of a city's morphological trajectory through time and, in turn, help planners collaboratively shape that trajectory. In tandem, spatial information technologies and urban morphology will further converge to generate new understandings of city pasts and presents and empower planners and community members in collaborative data-driven decision-making processes that center the human experience—physiological, psychological, and social—in the co-production of the urban form.

The interested reader is directed to the following resources to explore further:
- OpenStreetMap is available at https://www.openstreetmap.org/
- OSMnx is freely available at https://github.com/gboeing/osmnx
- Examples, tutorials, and demonstrations for using OSMnx (including how to reproduce the visualizations in this chapter) are available at https://github.com/gboeing/osmnx-examples

## Acknowledgments

Much of this chapter is adapted from an article published in the *International Journal of Information Management* (Boeing, 2019b). The author wishes to thank the publisher for permission to adapt it. Figures 5 and 6 are adapted from Boeing (2019a) under the terms of the Creative Commons Attribution 4.0 International License: http://creativecommons.org/licenses/by/4.0/